\documentclass[twocolumn,superscriptaddress,amsmath,aps,pra]{revtex4}

\usepackage[latin1]{inputenc}
\usepackage[T1]{fontenc}
\usepackage{amsmath,amssymb}
\usepackage{bbm}
\usepackage{graphicx}
\usepackage{subfigure}
\usepackage{sistyle}
\usepackage{ifpdf}
\usepackage{xcolor}

\DeclareMathOperator{\diag}{diag}

\def\be{\begin{equation}}
\def\ee{\end{equation}}
\def\bea{\begin{eqnarray}}
\def\eea{\end{eqnarray}}

\begin{document}

\title{Estimating a fluctuating magnetic field with a continuously monitored atomic ensemble}

\author{Cheng Zhang}
\affiliation{Department of Physics, Beijing Normal University, 100875, Beijing, China}
\affiliation{Department of Physics and Astronomy, University of Aarhus, DK-8000 Aarhus C, Denmark}

\author{Klaus M\o lmer}
\thanks{Corresponding author: moelmer@phys.au.dk}
\affiliation{Department of Physics and Astronomy, University of Aarhus, DK-8000 Aarhus C, Denmark}

\begin{abstract}

We study the problem of estimating a time dependent magnetic field by continuous optical probing of an atomic ensemble. The magnetic field is assumed to follow a stochastic Ornstein-Uhlenbeck process and it induces Larmor precession of the atomic ground state spin, which is read out by the Faraday polarization rotation of a laser field probe. The interactions and the measurement scheme are compatible with a hybrid quantum-classical Gaussian description of the unknown magnetic field, and the atomic and field variables. This casts the joint conditional quantum dynamics and classical parameter estimation problem in the form of update formulas for the first and second moments of the classical and quantum degrees of freedom. Our hybrid quantum-classical theory is equivalent with the classical theory of Kalman filtering and with the quantum theory of Gaussian states. By reference to the classical theory of smoothing and with the quantum theory of past quantum states, we show how optical probing after time $t$ improves our estimate of the value of the magnetic field at time $t$, and we present numerical simulations that analyze and explain the improvement over the conventional filtering approach.

\end{abstract}

\maketitle

\section{Introduction}

Quantum parameter estimation combines elements of classical estimation theory with quantum measurement theory to provide estimates of classical parameters or signals conditioned on the outcome of measurements on quantum systems \cite{maccone2004science}. In cases where probing is accomplished by measurements on a quantum probe, classical Kalman filter theory can thus be combined with the density matrix formalism to describe hybrid quantum-classical components, and yield the maximum likelihood estimator of classical variables.

Estimation of a weak classical magnetic field, is of both theoretical and practical interest in high-precision metrology. Atomic gases are excellent magnetic probes due to the Larmor precession of the atomic spin, which can be probed by a laser field \cite{romalis2003nature}.
The same probing, in turn, squeezes the collective atomic spin degree of freedom \cite{wiseman2001pra} and improves precision compared to the standard counting statistics limits of independent probe atoms.
The problem can be treated by Kalman filter theory \cite{maybeck1979,geremia2003}. For a recent, combined theoretical and experimental study, see \cite{morgan2018}. In \cite{stockton2004} and  \cite{molmer2004pra1}, a hybrid quantum-classical Gaussian-state formalism was proposed where the atomic and photonic degrees of freedom as well as an unknown constant magnetic field, were treated as harmonic oscillator quadrature variables. In the absence of atomic dissipation, the effect of atomic spin squeezing led to a $1/T^3$ rather than $1/T$ time dependence of the variance of the estimate of a constant magnetic field. Atomic dissipation can be included in the formalism \cite{molmer2004pra2} and limits the degree of squeezing  and prevents the long time $1/T^3$ resolution.

In \cite{stockton2004} and in  \cite{petersen2006pra}, the theory was generalized to the case of a magnetic field that fluctuates according to an Ornstein-Uhlenbeck process, and for which a hybrid Gaussian quantum-classical distribution still applies. This Gaussian distribution function is fully determined by the quadrature expectation values and covariances, which are conditioned on the interaction Hamiltonian and the measurement outcomes until time $t$. In \cite{petersen2006pra} it was speculated and proven that the use of measurement data acquired also after time $t$ could be employed to improve, in hindsight,  the estimate of the value of the magnetic field at time $t$.

For continuously monitored systems, {\it filtering} refers to the estimation of a classical signal at time $t$ conditioned upon observations until time $t$, while {\it smoothing} refers to estimation of the same quantity based on observations both before and after $t$. In classical estimation theory, smoothing is an integral part of Kalman filter theory \cite{maybeck1979,kalmansmoothing2016}. In \cite{tsang2009prl,tsang2009pra,tsang2010pra,tsang_phase_freq2008pra,tsang_2phase_freq2009pra}, Tsang showed how estimation by both classical filtering and smoothing can be generalized to the case of Gaussian quantum probes.  In the present article, we shall present an alternative derivation with starting point in the theory of quantum measurement theory, quantum trajectories and the past quantum state. The two approaches yield identical results when the systems are restricted to Gaussian phase space distributions, while our quantum approach may be readily adapted also to more general cases, which have their classical counterparts in the so-called forward-backward or $\alpha-\beta$ analysis of Hidden Markov Models \cite{scientific2007computing,herschlag2012online,molmer2014markovpra}.

The article is outlined as follows.
In Sec.~\ref{sec:mag}, we briefly describe the atomic magnetometer and show that its dynamics is captured by a few mode harmonic oscillator description.
In Sec.~\ref{sec:filtering}, we
present the Gaussian-state description of the unknown magnetic field and the collective quantum state of the atoms subject to optical probing.
In Sec.~\ref{sec:pqs},
we derive our main results, namely the Gaussian state mean values and covariance matrix for the filtering and smoothing analysis of the measurement record.
In Sec.~\ref{sec:results} we present numerical results of our scheme and address its performance in different limits.
Sec.~\ref{sec:dis} summarizes the paper and provides an outlook.

\section{An Atomic Ensemble Magnetometer}\label{sec:mag}
In this article we model a unidirectional time dependent magnetic field $B(t)$ by an Ornstein-Uhlenbeck process, governed by a stochastic equation
\begin{gather}
  \label{eq:OU}
  dB(t) = -\gamma_b B(t)dt + \sqrt{\sigma_{b}}dW,
\end{gather}
where $dW$ is an infinitesimal Wiener increment with mean value 0 and variance $dt$. An example of $B(t)$ is shown by the green curve in Fig. 1. The associated probability distribution for the unknown value of the magnetic field obeys a Fokker-Planck equation with constant friction and diffusion terms, and an initial Gaussian distribution will remain Gaussian for later times.

An ensemble of spin polarized atoms permits real-time tracking of an external time dependent magnetic field $B(t)$. We assume $N_{at}$ identical two-level atoms described by the Pauli spin matrices $\boldsymbol{\sigma}_i$ for the $i$th atom. We further assume the atomic gas is dilute so that scattering and interactions among the atoms are negligible. The atoms are prepared by optical pumping in the same internal quantum state, spin polarized along the $x$ direction. Mathematically, this allows us to treat the collective polarization along the $x$ axis classically $\langle J_x\rangle= N_{at}/2$, ($\hbar=1$), while the off-axis polarization components are quantum degrees of freedom and satisfy the commutation relation $[J_y,J_z]=i J_x$. This inspires us to introduce the effective canonical coordinate and momentum variables $x_\mathrm{at}=J_y/\sqrt{\langle J_x\rangle}, p_\mathrm{at}=J_z/\sqrt{\langle J_x\rangle}$ with the standard commutation relation $[x_\mathrm{at},p_\mathrm{at}]=i$.

In the presence of an external magnetic field $B$, polarized along the $y$ direction, the collective spin precesses toward the $z$ axis, which in terms of the canonical atomic variables corresponds to the time evolution
 \begin{eqnarray}
p_\text{at} \mapsto p_\text{at} - \mu\tau B,
\end{eqnarray}
during each infinitesimal time interval $\tau$ where $\mu=\beta\sqrt{\langle J_x\rangle}$ is given by the magnetic moment $\beta$.

In addition, the atoms interact with a continuous, linearly polarized (along $x$) beam of light, for which we adopt a simple description of the light-matter interaction by discretizing the beam into a sequence of segments of light with duration $\tau$. The Stokes operator for each segment of light with $N_\mathrm{ph}$ photons has an $x$ component which is effectively classical $\langle S_x\rangle= N_\mathrm{ph}/2$, while we may introduce the scaled canonical variables for the other two Stokes vector components, $x_\mathrm{ph}=S_y/\sqrt{\langle S_x\rangle}, p_\mathrm{ph}=S_z/\sqrt{\langle S_x\rangle}$, satisfying the commutation relation $[x_\mathrm{ph},p_\mathrm{ph}]=i$. Following \cite{molmer2004pra1}, the light-matter interaction in the Heisenberg picture is given by the two update rules
\begin{eqnarray}
x_\mathrm{at}\mapsto x_\mathrm{at}+\kappa\sqrt{\tau}p_\mathrm{ph}, && p_\mathrm{at}\mapsto p_\mathrm{at}, \\
x_\mathrm{ph}\mapsto \kappa\sqrt{\tau}p_\mathrm{at}+x_\mathrm{ph}, && p_\mathrm{ph}\mapsto p_\mathrm{ph},
\end{eqnarray}
where we introduce the coupling constant $\kappa =d^2\omega/{\Delta Ac\epsilon_0}\sqrt{N_\mathrm{at}N_\mathrm{ph}/\tau}$ with $d$ the atomic dipole moment, $\omega$ the photon frequency, $\Delta$ the detuning from atomic resonance, $A$ the area of the cross section of the light field and $\Phi =N_{ph}/\tau$ the photon flux \cite{julsgaard2006}. The dynamics of the atom and field variables is thus described by the effective Hamiltonian
\begin{eqnarray}\label{eq:ham}
H\tau = \kappa\sqrt\tau p_\mathrm{at} p_\mathrm{ph} + \mu\tau B x_\mathrm{at},
\end{eqnarray} 
where we recall that a new segment of light enters at each new time interval of duration $\tau$. Note that the atom-light interaction term in (\ref{eq:ham}) is of order $\sqrt\tau$ which ensures appropriate decoherence and noise properties associated with the detection of the light field after passage of the ensemble.

\section {Gaussian-state Formalism For Estimating a Time-Dependent-Noisy Magnetic Field}\label{sec:filtering}

The Hamiltonian (\ref{eq:ham}) and the Ornstein-Uhlenbeck process (\ref{eq:OU}) determine the evolution of the joint probability distribution of the magnetic field and the atomic and optical quadrature variables. This is accomplished by a density matrix $\rho(t)= \int dB |B\rangle\langle B|\otimes \rho_B(t)$, in which the different candidate values $B$ of the classical magnetic field are treated as if they were eigenvalues of a quantum observable with eigenstates populated in an incoherent manner, and the atomic spin and probe field occupy the unnormalized quantum states $\rho_B(t)$ which are correlated with the value of the $B$-field. The probability distribution of the magnetic field is then given by the expectation value of the projection operator $|B\rangle\langle B|$, and has the formal expression $P(B)=\mathrm{tr}(\rho_B(t))$.

We represent $\rho(t)$ by an effective hybrid classical-quantum Wigner function $W(\mathbf{y})$ with the five arguments $\mathbf{y} = (B,x_\mathrm{at}, p_\mathrm{at}, x_\mathrm{ph}, p_\mathrm{ph})^T$, for which the integral over all but one variable yields the marginal distribution for that variable. Due to the linear character of the problem, the Wigner function is Gaussian, and hence it is fully characterized by its mean values $\langle\mathbf{y}\rangle$ and its covariance matrix $\boldsymbol{\gamma}$  with elements
$\gamma_{ij} = 2 \text{Re} \langle(y_i-\langle y_i\rangle)(y_j-\langle y_j\rangle)\rangle$. We shall now recall the evolution of those quantities under the interactions and the measurement dynamics.

\subsection{Evolution of mean values and covariance matrix elements}

We assume that the time dependent magnetic field is not known by the observer, who will thus have recourse to a probabilistic description of the magnetic field. I.e., the Ornestein-Uhlenbeck process is not represented by a stochastic equation but by its effect on the first and second moments of the Gaussian probability distribution of the field and atomic variables.

Starting from the fully spin polarized state, the incident linearly polarized field, and prior Gaussian distribution for the magnetic field with zero mean and variance $V_b$, the joint Gaussian distribution is characterized by the vector of mean values and matrix of covariances
\begin{align}\label{eq:ini}
  \langle \mathbf{y}\rangle  &= (0,0,0,0,0),\\
  \boldsymbol{\gamma} &= \diag(2V_b,1,1,1,1).
  \end{align}
We partition the $5\times5$ covariance matrix $\boldsymbol{\gamma}$ and mean value vector $\langle \bold{y}\rangle$ into blocks
\begin{align}\label{eq:blockm}
  \boldsymbol{\gamma} &=
  \begin{pmatrix}
    \mathbf{A} & \mathbf{C}\\
    \mathbf{C}^T & \mathbf{B}
  \end{pmatrix}, \\
    \langle \mathbf{y^T}\rangle &= (\mathbf{m^T}, \mathbf{n^T}). \label{eq:blockv}
\end{align}
Here $\mathbf{m}$ and $\mathbf{A}$ denote the mean value vector and the $3\times3$ covariance matrix for the $B$-field and the atomic variables $(B,x_\text{at},p_\text{at})^T$, while $\mathbf{n}$  and $\mathbf{B}$ denote the mean value vector and the $2\times2$ covariance matrix for the field variables $(x_\text{ph},p_\text{ph})^T$, and $\mathbf{C}$ denotes the covariances between the optical field observables and the atoms and the $B$-field.

To describe the continuous probing, we note that each light segment approaching the atomic ensemble is not yet correlated with the atoms and the $B$ field, and hence we assume the values
  \begin{align}
  \mathbf{B} &\to \mathbbm{1}_{2\times2},\\
  \mathbf{C} &\to 0_{3\times2},\\
  \mathbf{n} &\to 0_{2\times1},
\end{align}
which we insert together with the current covariance matrix $\mathbf{A}$ and mean values $\mathbf{m}$ in (\ref{eq:blockm}) and (\ref{eq:blockv}), respectively.

The mean value of $B$ is damped by a rate $\gamma_b$, which affects also all covariance matrix elements involving $B$, and the variance of $B$ is furthermore subject to diffusive spreading with rate $\sigma_b$.
The Hamiltonian (\ref{eq:ham}) similarly causes a linear mixing of the mean values, and a corresponding mixing of the covariance matrix elements.

This leads to the following update rules during a small time interval $\tau$.
\begin{align}
  \langle\mathbf{y}\rangle &\to \mathbf{S}\langle\mathbf{y}\rangle,\\
  \boldsymbol{\gamma} &\to
  \mathbf{S}\boldsymbol{\gamma}\mathbf{S}^T+\bold{L}\label{eq:gammas},
\end{align}
where
\begin{gather}\label{eq:Sfor}
  \mathbf{S} =
  \begin{pmatrix}
    1-\gamma_b\tau & 0 & 0 & 0 & 0\\
    0 & 1 & 0 & 0 & \kappa \sqrt{\tau}\\
    -\mu \tau & 0 & 1 & 0 & 0\\
    0 & 0 & \kappa \sqrt{\tau} & 1 & 0\\
    0 & 0 & 0 & 0 & 1
  \end{pmatrix}
\end{gather}
and $\mathbf{L} = \diag(2\sigma_{b}\tau,0,0,0,0)$.

After the interaction, the atoms and the light segment have become correlated, which implies that the covariance matrix has nonvanishing entries between their corresponding components.

The light segment subsequently moves away from the atoms, and we may either disregard it, in which case the remaining components are merely described by a Gaussian distribution with mean values $\mathbf{m}$ and covarance matrix $\mathbf{A}$, or, we may perform a projective measurement of the photonic quadrature $x_\mathrm{ph}$ with a random measurement outcome $x_\mathrm{ms}$ (Gaussian distributed with variance $1/2$ \cite{current}). Due to the correlations in the Gaussian state, the projective measurement on the optical field updates the mean values and the covariances of the $B$-field and atomic variables according to

\begin{align}
\label{eq:A}
  \mathbf{A} &\to \mathbf{A}
  - \mathbf{C}_\gamma(\pi\mathbf{B}\pi)^-
  \mathbf{C}_\gamma^T, \\
   \mathbf{m} &\to \mathbf{m}
  + \mathbf{C}_\gamma (\pi\mathbf{B}_\gamma\pi)^-((x_\text{ms}-n_1),0)^T,  \label{eq:mA}
\end{align}
where $\pi=\diag(1,0)$ designates that we measure the first of the two light quadratures, and $(\dots)^-$ denotes the Moore-Penrose pseudoinverse. To lowest order $(\pi\mathbf{B}\pi)^-=\mathrm{diag}(1,0)$.

\begin{widetext}
As each segment of the light beam interacts only infinitesimally with the atomic system, the field variables can be eliminated, and to first order in the time increment $\tau$, the dynamics can be expressed in a closed set of equations for the mean values and the covariances of the $B$-field and atomic variables alone.  Denoting the matrix elements of $\mathbf{A}$ by $a_{ij}$, we get the deterministic update rule for $\mathbf{A}$,

\begin{align} \label{eq:reducedcov}
  \mathbf{A} \to
    \begin{pmatrix}
    (1-2\gamma_b\tau)a_{11}+2\sigma_b\tau & (1-\gamma_b\tau)a_{12} & (1-\gamma_b\tau)a_{13}-\mu\tau a_{11} \\
    (1-\gamma_b\tau)a_{21} & a_{22}+\kappa^2\tau & a_{23}-\mu\tau a_{21} \\
    (1-\gamma_b\tau)a_{31} -\mu \tau a_{11} & a_{32} - \mu\tau a_{12} & a_{33}-\mu\tau(a_{31}+a_{13})
  \end{pmatrix}
   -\kappa^2\tau
      \begin{pmatrix}
        a_{13}^2 & a_{13}a_{23} & a_{13}a_{33} \\
        a_{13}a_{23} & a_{23}^2 & a_{23}a_{33} \\
        a_{13}a_{33} & a_{23}a_{33} & a_{33}^2
  \end{pmatrix}.
  \end{align}
 \end{widetext}

The expected mean outcome of the field measurement is $\langle x_{ms}\rangle = n_1 = \kappa\sqrt{\tau}m_3$, and the  mean values for the magnetic field and atomic observables, conditioned on a given outcome $x_{ms}$ is given by
\begin{align} \label{eq:reducedmean}
  \mathbf{m} &\to
    \begin{pmatrix}
    (1-\gamma_b\tau)m_1 + \kappa\sqrt{\tau} a_{13} (x_{ms}-\kappa\sqrt{\tau} m_3)\\
    m_2 + \kappa\sqrt{\tau} a_{23} (x_{ms}-\kappa\sqrt{\tau} m_3)\\
    m_3 - \mu\tau m_1  + \kappa\sqrt{\tau} a_{33} (x_{ms}-\kappa\sqrt{\tau} m_3)
  \end{pmatrix}.
  \end{align}

Propagating these equations in subsequent steps of duration $\tau$, we acquire or simulate the detection record and determine the conditional dynamics of the atomic quantum state and the estimate of the current value of the magnetic field $B(t)$. The dynamics of the covariance matrix is deterministic and reaches a steady state irrespective of the measurement outcomes. The collective atomic spin and the B-field are correlated and the conditional probability distribution for $B$  has a Gaussian variance given by half of the first diagonal element of $\mathbf{A}$, around the first component of the vector of mean values, which in turn depends on the detection record.
\begin{figure}
	\centering
 		\label{subfig:estforB}
		\begin{minipage}[b]{0.5\textwidth}
			\centering
			\includegraphics[clip, width=0.9\textwidth]{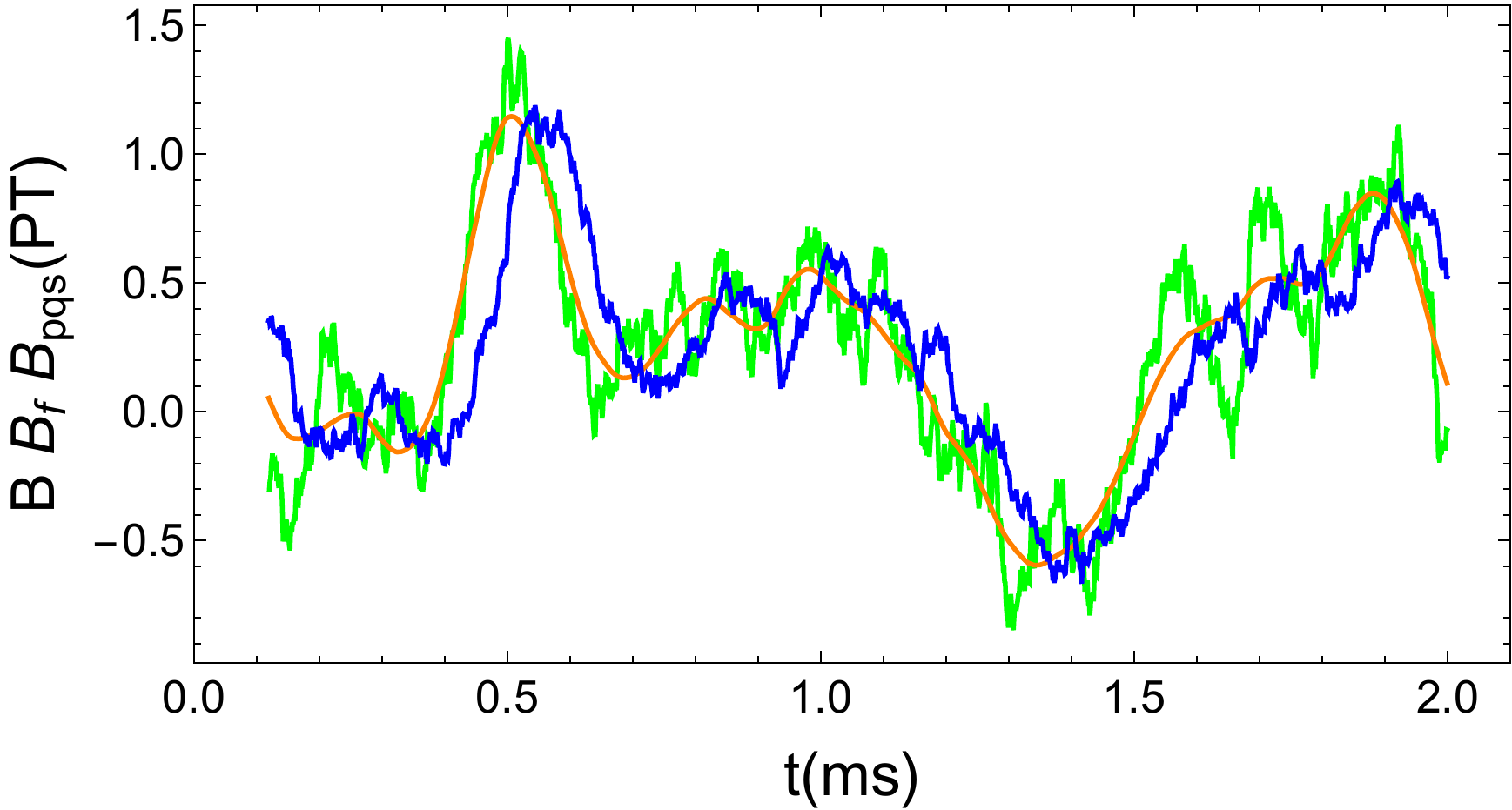}
 		\end{minipage}

    \caption{The green curve shows a simulated field $B(t)$, fluctuating according to the Ornstein Uhlenbeck process (\ref{eq:OU}) with parameters $\gamma_b=1.0\times10^{3}s^{-1}$, $\sigma_b=1.0\times10^{3}pT^2/s$. The field is monitored by the atomic ensemble coupled with strengths $\mu=2.0\times{10^5}s^{-1}$ to the magnetic field and $\kappa^2={10^4}s^{-1}$ to the optical probe. The simulated detection record
    yields the blue curve $B_f(t)$ by conventional forward filtering and the orange curve $B_\mathrm{PQS}(t)$ by the PQS (smoothing) scheme.
		}
    \label{fig:estB}
\end{figure}

In Fig. 1 we show a simulated realization of the noisy $B(t)$ (green curve). The blue curve shows our estimated $B(t)$ by the procedure outlined above. We observe an overall good agreement, but we also note that individual spikes in $B(t)$ are not reproduced, while other spikes appear. This reflects that the data acquisition is not fast enough to resolve rapid changes of $B(t)$, while the measurement shot noise may cause erroneous variations in the magnetic field estimate. Notably, the time dependence of the blue curve lags behind the green one. This is because changes in $B(t)$ are accumulated over time in the value of the spin precession angle and are only reliably discerned after a suitable optical signal has been obtained.

\section{Retrodiction of the magnetic field and atomic state}\label{sec:pqs}

In the previous section, we determined the joint Gaussian probability distribution of the magnetic field and the atomic collective spin at time $t$, conditioned on the probing data obtained until time $t$.
In the quantum theory of measurements, each interaction and probing event with outcome $m_i$ is formally described by a POVM element, and the full optical detection process of our scheme is described by applying a sequence of operators $M_{m_i}$, including both the deterministic time evolution of the state and the evolution conditioned on the measurement outcomes $m_i $. The joint probability for the occurrence of the full sequence of measurement outcomes is the trace of the corresponding operator product, $P(m_1,m_2, \ldots  m_N) =\mathrm{tr}(M_{m_N} \dots M_{m_1}\rho(0)M^{\dagger}_{m_1} \dots  M^{\dagger}_{m_N})$, while the joint probability of all measurement outcomes {\it and} a projective measurement of the magnetic field yielding the value $B$ at an intermediate time $t$ reads, $P(m_1,m_2, \dots  B_t, \ldots   m_N) =\mathrm{tr} (M_{m_N} \dots |B\rangle\langle B| \dots   M_{m_1}\rho(0)M^{\dagger}_{m_1} \dots  |B\rangle\langle B| \dots M^{\dagger}_{m_N})$.

Note that if the optical probing stops just before time $t$, the conditional quantum state at this time reads $\rho(t) \propto  M_{m_N} \dots  M_{m_1}\rho(0)M^{\dagger}_{m_1} \dots M^{\dagger}_{m_N}$ and the inferred conditional probability agrees with the conventional Born rule, $P(B) = \mathrm{tr}(|B\rangle\langle B|\rho(t))$. In the case of continued probing {\em after} $t$, we can use the cyclic property of the trace and reorganize terms to write the joint probability distributions as $P(m_1,m_2, \ldots B_t, \ldots   m_N) =\mathrm{tr}(|B\rangle\langle B|( \ldots   M_{m_1}\rho(0)M^{\dagger}_{m_1} \ldots  )|B\rangle\langle B|( \ldots   M^{\dagger}_{m_N}M_{m_N} \dots ))$.

Since the outcomes $m_i$ are the ones actually measured, we infer the probability that the magnetic field would have been projectively measured to have the value $B$, conditioned on all prior and posterior detection events to be proportional to
$\mathrm{tr} (|B\rangle\langle B|\rho(t)|B\rangle\langle B| E(t))$\cite{molmer2015prl}, with $\rho(t)= ( \dots   M_{m_1}\rho(0)M^{\dagger}_{m_1} \ldots )$ and $E(t)=( \dots   M^{\dagger}_{m_N}M_{m_N} \dots )$ where the $ \dots  $ in the expressions for $\rho(t)$ and $E(t)$ represent the sequences of POVM elements, until and after the time $t$, respectively.

\subsection{Backward evolution of the effect operator}

The POVM operators for optical probing can be determined as integrals with Gaussian kernels, see, e.g., \cite{xiao2020nature}, but we shall have recourse to a simplified argument that utilizes the more convenient representation of the operators $\rho(t)$ and $E(t)$ in tems of Gaussian mean values and covariance matrices. This representation applies because both operators are evolved by Gaussian preserving elements and hence both have Gaussian Wigner functions (any operator on an effective position and momentum operator phase space has a Wigner function representation, and being a hermitian and positive operator, $E(t)$ has a Wigner function with similar properties as the one of a conventional quantum state $\rho(t)$. Indeed, the time evolution of the $\rho$ and the $E$ operators are equivalent except that $E(t)$ evolves from the later towards the earlier times. As a consequence, $E(t)$ is described by a Gaussian Wigner function, and its first and second moments evolve with similar factors as the moments for $\rho(t)$.

We describe $E(t$) by a covariance matrix $\boldsymbol{\gamma}_E(t)$ and a vector of mean values $\langle  \mathbf{y}_E \rangle = (\mathbf{m}_E,\mathbf{n}_E)$, for which the evolution from $t+\tau$ to $t$ yields,
\begin{align}
  \langle \mathbf{y}_E \rangle  &\to \mathbf{S}_E  \langle \mathbf{y}_E \rangle, \\
  \boldsymbol{\gamma}_E &\to
  \mathbf{S}_E\boldsymbol{\gamma}_E\mathbf{S}_E^T+\bold{L}\label{eq:gammas},
\end{align}
where
\begin{gather}
  \label{eq:9}
  \mathbf{S}_E =
  \begin{pmatrix}
    1+\gamma_b\tau & 0 & 0 & 0 & 0\\
    0 & 1 & 0 & 0 & -\kappa \sqrt{\tau}\\
    \mu \tau & 0 & 1 & 0 & 0\\
    0 & 0 & -\kappa \sqrt{\tau} & 1 & 0\\
    0 & 0 & 0 & 0 & 1
  \end{pmatrix},
\end{gather} \\
and $\mathbf{L} =\diag(2\sigma_{b}\tau,0,0,0,0)$.

\begin{widetext}
Going through the matrix multiplications and eliminating the optical field components, we obtain the deterministic evolution of the magnetic field and atomic components covariance matrix, backward in time from $t+\tau$ to $t$ for the operator $E(t)$.

Denoting the matrix elements of $\mathbf{A}_E$ by $a_{ij}$, we get the deterministic update rule for $\mathbf{A}_E$,
\begin{align} \label{eq:reducedcovE}
\mathbf{A}_E  \to
   \begin{pmatrix}
    (1+2\gamma_b\tau)a_{11}+ 2\sigma_b\tau & (1+\gamma_b\tau)a_{12} & (1+\gamma_b\tau)a_{13}+\mu\tau a_{11} \\
    (1+\gamma_b\tau)a_{21} & a_{22}+\kappa^2\tau & a_{23}+\mu\tau a_{21} \\
    (1+\gamma_b\tau)a_{31} +\mu \tau a_{11} & a_{32} + \mu\tau a_{22} & a_{33}+ \mu\tau(a_{31}+a_{13})
  \end{pmatrix}
  -\kappa^2\tau
      \begin{pmatrix}
        a_{13}^2 & a_{13}a_{23} & a_{13}a_{33} \\
        a_{13}a_{23} & a_{23}^2 & a_{23}a_{33} \\
        a_{13}a_{33} & a_{23}a_{33} & a_{33}^2
  \end{pmatrix}.
  \end{align}
\end{widetext}
The centroid of the Gaussian Wigner function for $E(t)$ depends on the  measurement outcome $x_{ms}$ and is given by

\begin{align} \label{eq:reducedmeanE}
  \mathbf{m}_E &\to
    \begin{pmatrix}
    (1+\gamma_b\tau)m_1 + \kappa\sqrt{\tau} a_{13} (x_{ms}+\kappa\sqrt{\tau} m_3)\\
    m_2 + \kappa\sqrt{\tau} a_{23} (x_{ms}+\kappa\sqrt{\tau} m_3)\\
    m_3 + \mu\tau m_1  + \kappa\sqrt{\tau} a_{33} (x_{ms}+\kappa\sqrt{\tau} m_3)
  \end{pmatrix},
  \end{align}
where we emphasize that in these equations, the matrix and vector elements $a_{ij}$, $m_i$ are the ones pertaining to $\mathbf{A}_E$ and $\mathbf{m}_E$ and not to  $\mathbf{A}_{\rho} \equiv \mathbf{A}$ and $\mathbf{m}_{\rho}\equiv \mathbf{m}$ in (\ref{eq:reducedcov},\ref{eq:reducedmean}).

\subsection{Past quantum state estimate of the time dependent magnetic field}

Given the first and second moments of Gaussian states, we have the information needed to construct the $\rho(t)$ and $E(t)$ matrices, and may thus, in principle, determine the probability distribution for the outcomes of any measurement process at time $t$ conditioned on all previous and later measurements. The evaluation of arbitrary operator expressions from Wigner functions is, however, not a trivial one, and to avoid a cumbersome translation between operator expressions and their Wigner function phase space equivalents, we shall employ arguments to arrive at the desired result directly in terms of the conditional mean values and covariance matrices.

First, we note that $\mathrm{tr_B}(|B\rangle\langle B|\rho)= \langle B|\rho|B\rangle$ and $\mathrm{tr_B}(|B\rangle\langle B|E)=\langle B|E|B\rangle$, where $\mathrm{tr_B}$ denotes the partial trace over the $B$ degree of freeedom, are operators on the atomic spin Hilbert space. This leads to the observation that
\begin{align}
\label{}
   P_\mathrm{PQS}(B) &\propto \mathrm{tr}(|B\rangle\langle B|\rho|B\rangle\langle B|E) \nonumber \\
                     &\propto \mathrm{tr}_r \langle B|\rho|B\rangle\langle B|E|B\rangle,
\end{align}
where $\mathrm{tr}_r$ denotes the reduced trace over the atomic collective spin (oscillator) variables. The trace of a product of operators (a scalar product on the space of operators) equals $2\pi$ times the phase space integral of the product of the corresponding Wigner functions for a harmonic oscillator system. Hence, we obtain
\begin{align}
\label{}
   P_\mathrm{PQS}(B) &\propto \int dx_\mathrm{at} dp_\mathrm{at} W_{\rho}(B,x_\mathrm{at},p_\mathrm{at}) W_{E}(B,x_\mathrm{at},p_\mathrm{at}).
\end{align}
We now use the fact that the Wigner functions are Gaussian distributions and write their product explicitly as
\begin{align}
\label{eq:Wpqs}
    \Pi_{\rho,E}(B,x_\mathrm{at},p_\mathrm{at})&=W_{\rho}(B,x_\mathrm{at},p_\mathrm{at}) W_{E}(B,x_\mathrm{at},p_\mathrm{at})  \nonumber \\    			
    &\propto e^{-(\mathbf{y}-\mathbf{m}_{\rho})^{T}\mathbf{A}_{\rho}^{-1}(\mathbf{y}-\mathbf{m}_{\rho})}
    e^{-(\mathbf{y}-\mathbf{m}_{E})^{T}\mathbf{A}_{E}^{-1}(\mathbf{y}-\mathbf{m}_{E})}						
\end{align}
where $\mathbf{y} = (B,x_{at},p_{at})^{^T}$, $\mathbf{m}_{\rho(E)}$ denote the displaced mean values, and $\mathbf{A}_{\rho(E)}$ the $3\times 3$ covariance matrices of the magnetic field and atomic spin components of $\mathbf{y}$, in the Gaussian distributions $W_{\rho(E)}$. After elementary algebra, we can rewrite the product $\Pi_{\rho,E}$ in a single Gaussian form
\begin{equation}
\label{ }
\Pi_{\rho,E}\propto e^{-(\mathbf{y}-\mathbf{m}_{\rho,E})^{T}\mathbf{A}_{\rho,E}^{-1}(\mathbf{y}-\mathbf{m}_{\rho,E})}
\end{equation}
with the new covariance matrix $\mathbf{A}_{\rho,E}$ and mean value $\mathbf{m}_{\rho,E}$ given by
\bea
\mathbf{A}_{\rho,E}^{-1} &=& \mathbf{A}_{\rho}^{-1} + \mathbf{A}_{E}^{-1} ,\\
\mathbf{m}_{\rho,E}&=& \mathbf{A}_{\rho,E} (\mathbf{A}_{\rho}^{-1} \mathbf{m}_{\rho}+\mathbf{A}_E^{-1}\mathbf{m}_{E} ).
\eea
Therefore in the PQS framework the estimated magnetic field and variance of its conditional distribution are given by the first vector component and matrix elements
\begin{align}
     \label{eq:Bpqs}
     &B_\mathrm{PQS}=[(\mathbf{A}_{\rho}^{-1} + \mathbf{A}_{E}^{-1})^{-1}(\mathbf{A}_{\rho}^{-1} \mathbf{m}_{\rho}+\mathbf{A}_E^{-1}\mathbf{m}_{E})]_1
     ,    \\
     \label{eq:varpqs}
     &\mathrm{Var}_\mathrm{PQS}(B)=\frac{1}{2}[(\mathbf{A}_{\rho}^{-1} + \mathbf{A}_{E}^{-1})^{-1}]_{11}.
\end{align}

Eqs.(\ref{eq:Bpqs},\ref{eq:varpqs}) together with the equations to determine their constituents are the main results of this article. Given the probing record, they yield a Bayesian estimate of the time dependent magnetic field strength in form of a Gaussian distribution. In the next section we shall address the performance of the estimation and, in particular the difference between usual filter estimation and the Past Quantum State estimation.

\section{Results}\label{sec:results}

We have applied the PQS formalism to
the estimation of a time-dependent stochastic
magnetic field using an atomic ensemble and a light field as probes.
In Fig.~\ref{fig:estB} we show a simulated magnetic field $B(t)$ generated by the OU process (\ref{eq:OU}) (green curve) and the results
of the the quantum filtering and PQS estimation schemes. It is clear
that the estimates $B_f, B_\mathrm{PQS}$ follow the real field $B(t)$ approximately
for both methods. However, for the quantum filtering scheme (blue line) we observe a time
lag between the estimated field and its real value, whereas no visible
lag can be observed for the PQS scheme (orange line).
We also observe that $B_\mathrm{PQS}$ is smoother than $B_f$. While causing back action on the mean value vectors $\mathbf{m}_\rho$ and  $\mathbf{m}_E$, as time passes from $t$ to $t+\tau$, the contribution of the photon shot noise in the intervening interval merely passes from $E(t+\tau)$ to $\rho(t+\tau)$, which suppresses its effect on the estimate (\ref{eq:Bpqs}). Neither $B_f(t)$ nor $B_{PQS}(t)$ are capabale of resolving rapid fluctuations in the true magnetic field $B(t)$.

In parameter estimation, a major measure is the precision of the estimator. Following the literature we have tacitly assumed that the width of the forward and PQS Gaussian probability distributions are indicators of the precision of the estimate. Indeed, for a given record theses widths do provide the Bayesian probability distribution of the actual value, but this is not the same as the distribution over many runs of the difference between the inferred, maximum likelihood value (the mean value of the conditional Gaussian distribution) and the true value.

\begin{figure}
	\centering
		\includegraphics[width=0.48\textwidth]{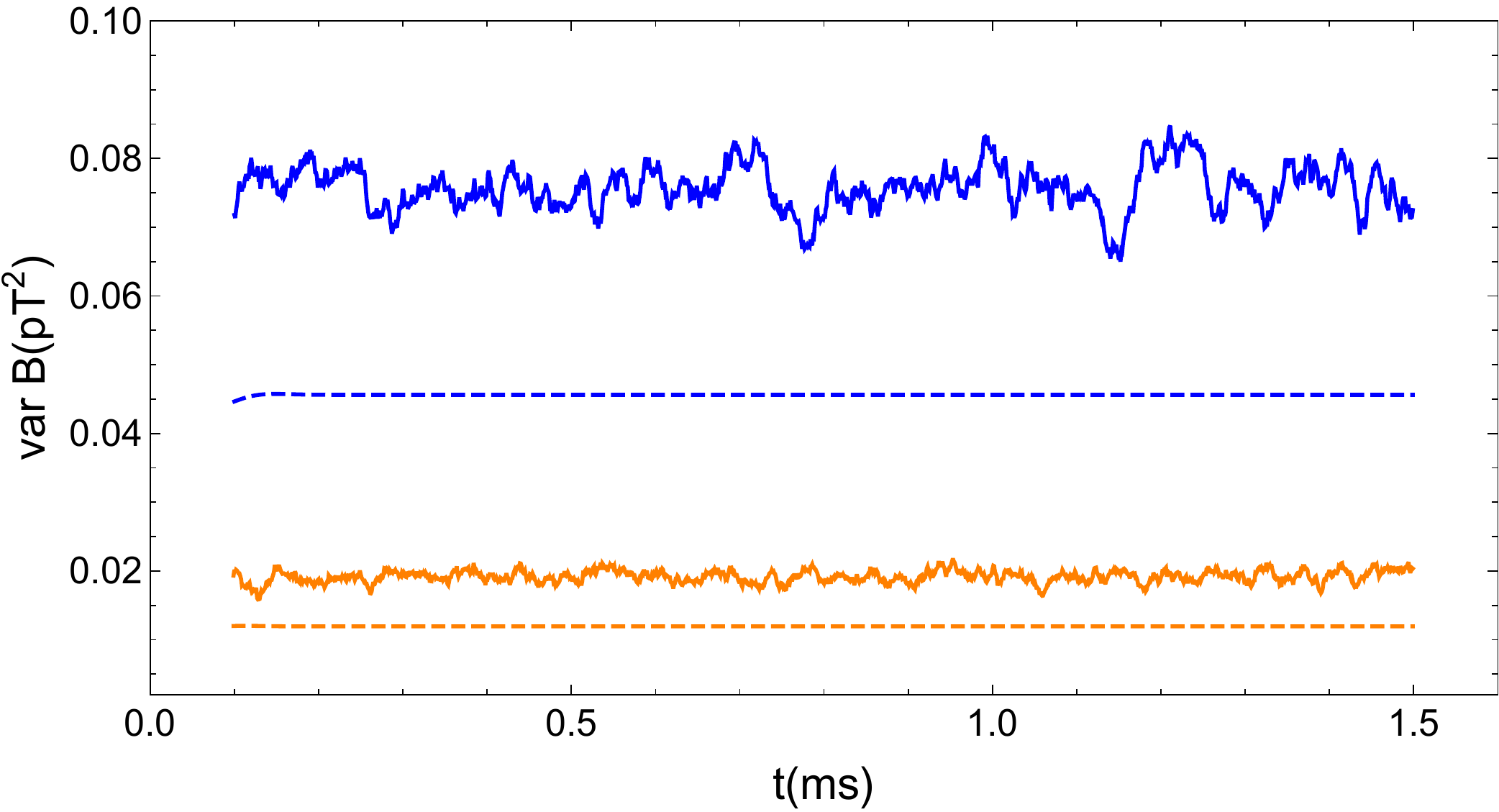}
	\caption{Time dependence of the variance of the estimated
		      magnetic field $B_f(t),B_\mathrm{pqs}(t)$. Parameters
		      we used here are identical to those in Fig. \ref{fig:estB}.
		      The dashed blue and orange lines indicate the variances
		      of the conditional Gaussian B-distributions  for the forward
		      quantum filter and the PQS analysis, respectively.
		      The solid blue and orange curves show the squared deviation
		      of the maximum likelihood estimator (33) from the true simulated
		      value. The error of PQS scheme is reduced by an approximate
		      factor of 4 compared to quantum filtering.
		      }
	\label{fig:var}
\end{figure}

To characterize the error of our estimator, we evaluate the mean squared error (MSE) over
an ensemble of $M=2000$ independent realizations of the real magnetic field $B(t)$,
which is defined by
\begin{equation}
\label{eq:mse}
	var B(t) =\frac{1}{M}\sum_{k=1}^M(B_{\mathrm{est},k}(t)-B_k(t))^2,
\end{equation}
where $B_k(t)$ is the $k$th realization of the
magnetic field and $B_{\mathrm{est},k}(t)$ is the corresponding estimated
result with our filter and PQS schemes. Fig. \ref{fig:var},
displays the numerically calculated variances from (\ref{eq:mse})  and the variances of the Gaussian distributions (\ref{eq:A}) and (\ref{eq:varpqs}). We see that the variance by the PQS scheme is approximately 4 times smaller than by the filter scheme, whether we compare the Bayesian widths of every individual realization (dashed curves) or the statistical deviation of the maximum likelihood estimate from the true value (solid curves).

This is an interesting result. Due to the doubling of experimental data available to the PQS field estimate at time $t$ (probing both before and after $t$), one might have expected an approximate factor of two improvement, cf., the sum of the inverse $\rho$ and $E$ covariance matrices in Eq.(\ref{eq:varpqs}). This expression, however, involves the full covariance matrices and not only the $a_{11}$-components representing the variances of the $B$-field. Indeed, $\mathrm{Var}_\mathrm{PQS}(B) \ne \frac{1}{2} ((\mathbf{A}_{\rho}^{-1})_{11} + (\mathbf{A}_{E}^{-1})_{11})^{-1}$. The fact that our $B$-field estimate is correlated with the unmeasured atomic spin components, and these components appear with the same arguments inside both Gaussian functions $W_\rho$ and $W_E$, narrows the range of $B$ values by another factor of two.

A second observation is the slightly less than factor two disagreement between the variance of the individually filtered or smoothened $B$-distributions and the variance of the deviation between the maximum likelihood estimator and the true value. Our calculations show that the Bayesian and maximum likelihood estimator yield identical variances for the estimation of a constant magnetic field. The best estimator is known to have fluctuations which are lower bounded by the reciprocal of the Fisher information $I$. More elaborate methods exist to calculate the Fisher information for continuous probing \cite{Gammelmark}, and for a single variable, the Bayesian filter estimator is known to reach that limit asymptotically. We are interested in estimating a varying  magnetic field, and we should hence evaluate a corresponding {\it Fisher information matrix} $I_{ij}$ for the field at times $t_i$. The methods successfully applied in  \cite{Genoni2017,Genoni2017njp,Genoni2018quatum} to obtain the Fisher information for the case of continuous monitoring with Gaussian states and variables may be a good starting point for the calculation of such a multi-time Fisher information matrix for the time varying field.

\begin{figure}
	 \subfigure{}{
 		\label{subfig:var_gam}
		\begin{minipage}[b]{0.48\textwidth}
			\centering
			\includegraphics[clip, width=0.9\textwidth]{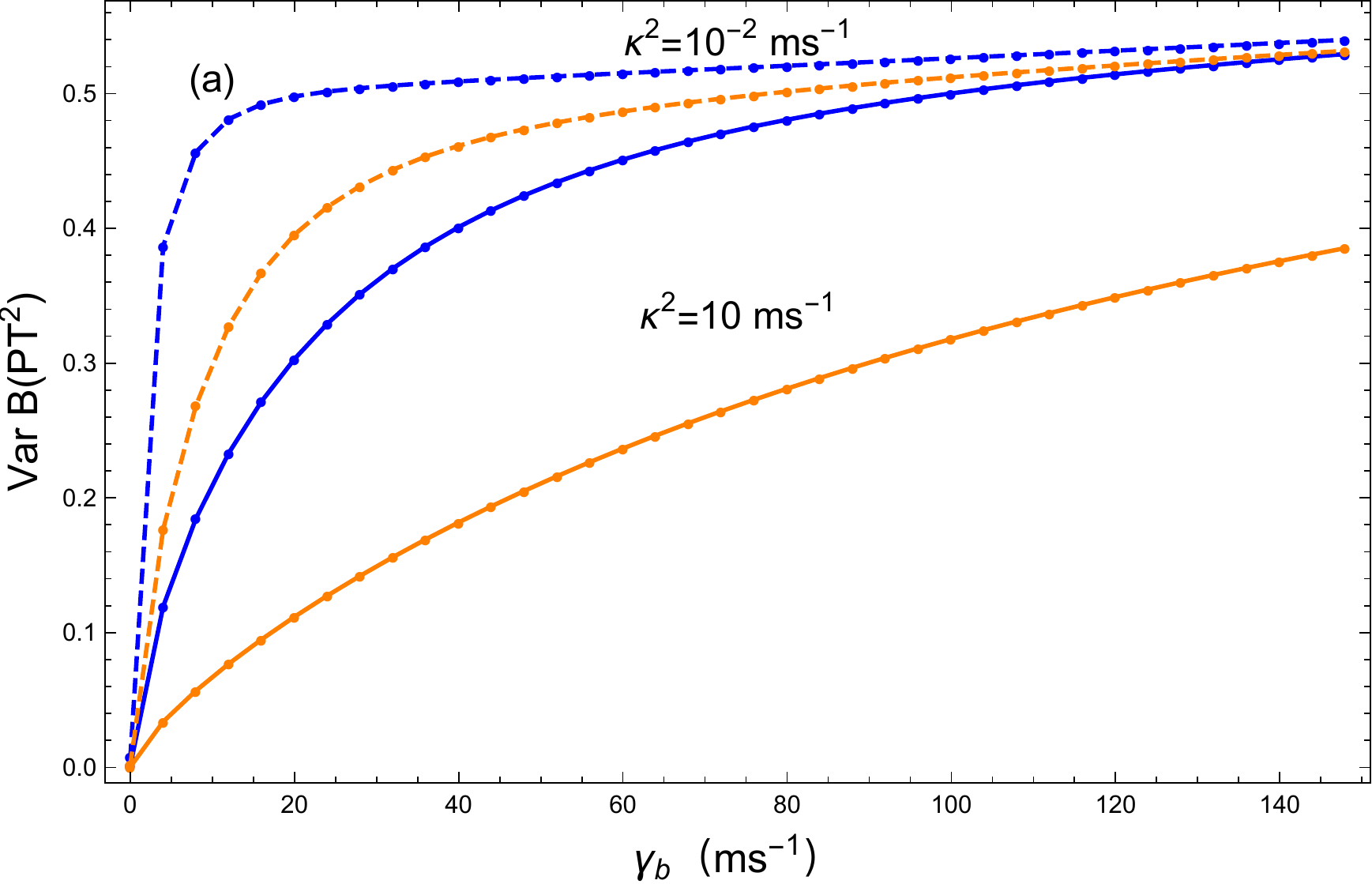}
 		\end{minipage}}

	\subfigure{}{
 		\label{subfig:var_kap}
		\begin{minipage}[b]{0.48\textwidth}
			\centering
			\includegraphics[clip, width=0.9\textwidth]{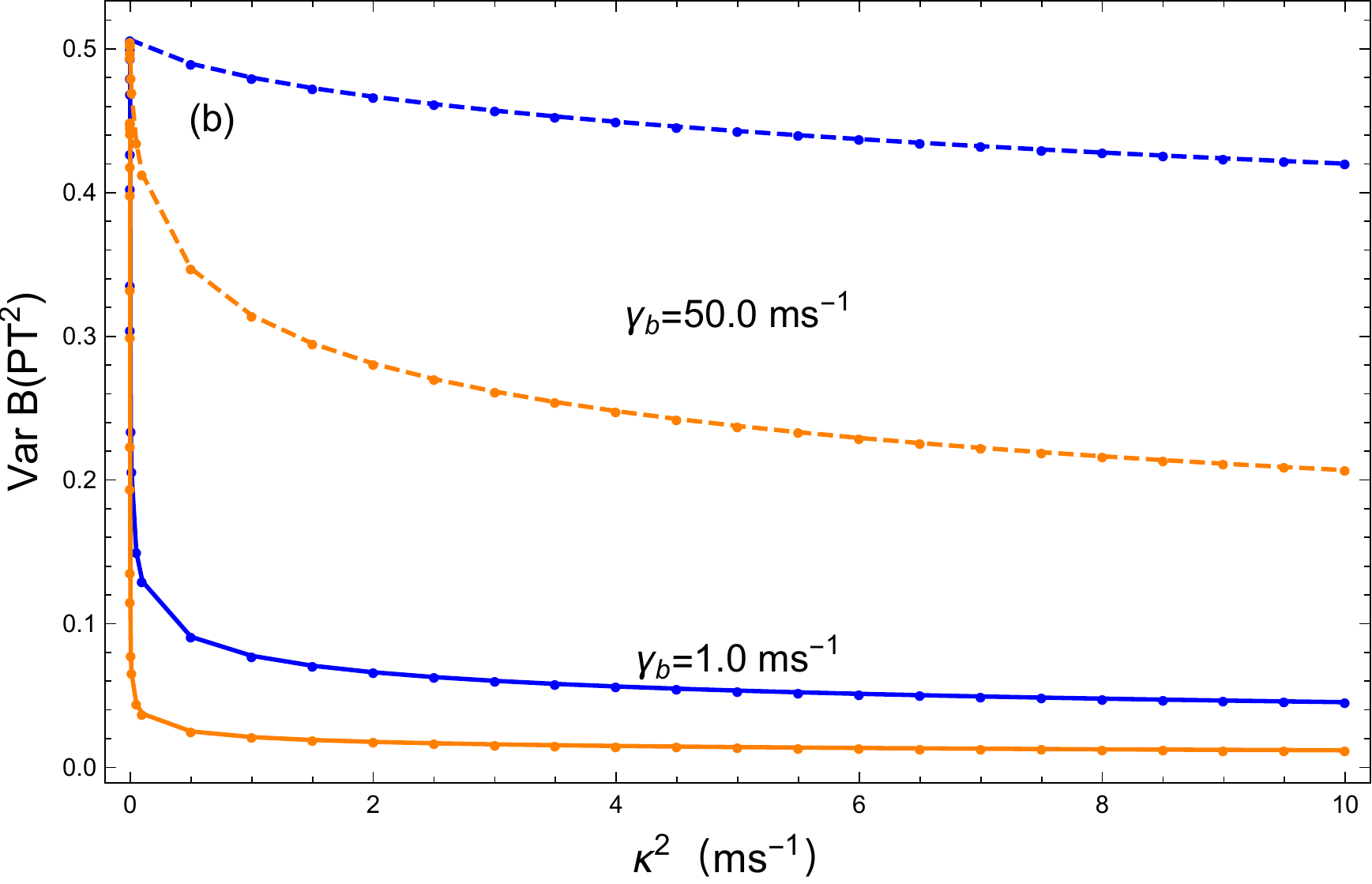}
 		\end{minipage}}

    \caption{The theoretically predicated conditional variances of $B(t)$ in steady state as a function of $\gamma_b$
    	          and $\kappa^{2}$ in Fig. 3 (a) and (b) respectively. The blue curves represent filtering process while the
	          orange curves describe the PQS scheme. Dots indicate numerical data and lines are guides to the eyes. (a) Plot of the conditional variance of $B(t)$ as a function of
	          $\gamma_b$ for a fixed value of $\kappa^2=10^{-2}ms^{-1} $ (dashed blue and orange lines) and $\kappa^2=10ms^{-1}$ (solid blue and orange lines) . (b)Conditional variance of $B(t)$ as a function of  $\kappa^2$ for a fixed value of $\gamma_b=1.0ms^{-1}$ (dashed blue and orange lines) and $\gamma_b=50.0ms^{-1}$ (solid blue and orange lines). We fix the ratio $\sigma_b/\gamma_b=1pT^2$
	          and $\mu=200ms^{-1}$ here.}
    \label{fig:var_para}
\end{figure}

Finally, let us note that the advantage of the PQS scheme changes under variation of parameters, and for example a very rapidly fluctuating magnetic field may neither be estimated by the filter nor the PQS theory, and its value is ultimately only restricted by the steady state distribution of
the (unobserved) process with variance $\sigma_b/2\gamma_b$.  To illustrate this, we show
in Fig.~\ref{fig:var_para}
the theoretically predicated variances  of the estimated fields by the filter and PQS analyses
as functions the Ornstein-Uhlenbeck rate parameter $\gamma_b$ and
the probing strength $\kappa^2$. With a given, finite probing strength $\kappa^2$, the time evolution of the magnetic field is tracked progressively better by both methods when $\gamma_b < \kappa^2$. And it is in the same regime that the PQS advantage over forward filtering is maximal.

\section{Discussion}\label{sec:dis}
In this paper, we have developed a theory for the estimation
of a time-dependent magnetic field generated by an Ornstein-Uhlenbeck process
based on the quantum filtering and PQS schemes. Our
numerical results confirm and explain an enhanced precision of the estimate of the time dependent magnetic field by full measurement records over the conventional quantum filtering approach.
Our hybrid quantum-classical theory is equivalent with the classical theory of Kalman filtering and smoothing on the one hand, and with the quantum theory of quantum trajectories and past quantum states on the other hand. We believe that the combined insight from these two domains of precision metrology may play a crucial role and may point to the use of further theoretical methods in the very active field of high precision measurements and hypothesis testing with quantum systems.

\begin{acknowledgments}
  The authors acknowledge support from the Villum Foundation
  and the Chinese Scholarship Council (CSC).
\end{acknowledgments}

\setcounter{equation}{0}
\setcounter{figure}{0}

%



\begin{thebibliography}{10}
\bibitem{current}
In the analysis of a real experimental record, $x_\mathrm{ms}$ is the actual measurement outcome, while to simulate an expermiment we evolve only the atomic and photonic components of Eqs.~(\ref{eq:ini})-(\ref{eq:mA}) subject a given stochastic realization of $B(t)$ according to the OU process. We then choose $x_\mathrm{ms}$ at random at each time step, according to a Gaussian distribution with mean value $n_1$ and variance $1/2$.
\end{thebibliography}
\end{document}